\PassOptionsToPackage{table}{xcolor}
\documentclass[sigconf, screen]{acmart}
\usepackage[nolist]{acronym}
\usepackage{enumitem}
\usepackage{graphicx}
\usepackage{hyperref}
\usepackage{makecell}
\usepackage{siunitx}
\usepackage{subcaption}

\definecolor{pastelorange}{HTML}{FFE5CC} 
\definecolor{pastelblue}{HTML}{D4E1F5}   
\definecolor{pastelgreen}{HTML}{D4F5D4}  

\definecolor{pastelpurple}{HTML}{9467BD} 
\definecolor{pastelcyan}{HTML}{17bECF}   
\definecolor{pastelred}{HTML}{D62728}  

\def\wthree{0.325\textwidth}

\AtBeginDocument{%
  }

\copyrightyear{2026}
\acmYear{2026}
\setcopyright{cc}
\setcctype{by}
\acmConference[CHI EA '26]{Extended Abstracts of the 2026 CHI Conference on Human Factors in Computing Systems}{April 13--17, 2026}{Barcelona, Spain}
\acmBooktitle{Extended Abstracts of the 2026 CHI Conference on Human Factors in Computing Systems (CHI EA '26), April 13--17, 2026, Barcelona, Spain}
\acmDOI{10.1145/3772363.3798498}
\acmISBN{979-8-4007-2281-3/2026/04}

\begin{document}

\title{CR-Eyes: A Computational Rational Model of Visual Sampling Behavior in Atari Games}

\author{Martin Lorenz}
\email{lorenz@cs.uni-leipzig.de}
\orcid{0009-0003-3517-5839}
\affiliation{%
  \institution{ScaDS.AI, Leipzig University}
  \city{Leipzig}
  \country{Germany}
}

\author{Niko Konzack}
\email{be66akax@studserv.uni-leipzig.de}
\orcid{0009-0001-6966-2497}
\affiliation{%
  \institution{Leipzig University}
  \city{Leipzig}
  \country{Germany}
}

\author{Alexander Lingler}
\email{alexander.lingler@it-u.at}
\orcid{0009-0004-9439-7375}
\affiliation{%
  \institution{Interdisciplinary Transformation University}
  \city{Linz}
  \country{Austria}
}

\author{Philipp Wintersberger}
\email{philipp.wintersberger@it-u.at}
\orcid{0000-0001-9287-3770}
\affiliation{%
  \institution{Interdisciplinary Transformation University}
  \city{Linz}
  \country{Austria}
}

\author{Patrick Ebel}
\email{ebel@uni-leipzig.de}
\orcid{0000-0002-4437-2821}
\affiliation{%
  \institution{ScaDS.AI, Leipzig University}
  \city{Leipzig}
  \country{Germany}}

\renewcommand{\shortauthors}{Lorenz et al.}

\begin{abstract}

Designing mobile and interactive technologies requires understanding how users sample dynamic environments to acquire information and make decisions under time pressure. 
However, existing computational user models either rely on hand-crafted task representations or are limited to static or non-interactive visual inputs, restricting their applicability to realistic, pixel-based environments.
We present CR-Eyes, a computationally rational model that simulates visual sampling and gameplay behavior in Atari games. Trained via reinforcement learning, CR-Eyes operates under perceptual and cognitive constraints and jointly learns where to look and how to act in a time-sensitive setting.
By explicitly closing the perception–action loop, the model treats eye movements as goal-directed actions rather than as isolated saliency predictions.
Our evaluation shows strong alignment with human data in task performance and aggregate saliency patterns, while also revealing systematic differences in scanpaths.
CR-Eyes is a step toward scalable, theory-grounded user models that support design and evaluation of interactive systems.

\end{abstract}


\begin{acronym}[]
    \acro{ALE}{Arcade Learning Environment}
    \acro{AOI}{Area of Interest}
    \acro{CR}{Computational Rationality}
    \acro{HCI}{Human-Computer Interaction}
    \acro{GOMS}{Goals, Operators, Methods, and Selection rules}
    \acro{KLM}{Keystroke-Level Model}
    \acro{MDP}{Markov Decision Process}
    \acro{OS}{Operating System}
    \acro{POMDP}{Partially Observable Markov Decision Process}
    \acro{PPO}{Proximal Policy Optimization}
    \acro{PVM}{Persistence-of-Vision Memory}
    \acro{RL}{Reinforcement Learning}
    \acro{SA}{Situation Awareness}
    \acro{SUGARL}{Sensorimotor Understanding Guided Active Reinforcement Learning}
    \acro{UCD}{User-Centered Design}
    \acro{UI}{User Interface}
    \acro{UX}{User Experience}
    \acro{UXD}{User Experience Design}
    \acro{XAI}{Explainable AI}
\end{acronym}


\maketitle

\section{Introduction}

Understanding human perception during visually complex, dynamic tasks is essential for designing technologies that people can use efficiently in everyday life. Whether glancing at a smartwatch while navigating a crowded street or making rapid decisions in fast-paced video games, users must continuously determine where to direct attention and how to act accordingly. These decisions are made under perceptual and cognitive constraints with the goal to maximize task performance or expected utility~\cite{gershmanComputationalRationalityConverging2015}. This fundamental need for efficient information acquisition under time-pressure links gaming to high-stakes real-world domains, despite differing costs of failure.
The ability to predict scanpaths can inform the design of intelligent user interfaces, adaptive notification systems, and assistive technologies~\cite{gasse_evaluating_2025, lingler2024Supporting}.
Models that close the perception-action loop by treating gaze as a goal-directed action that determines what users perceive next, and how this affects their behavior can serve as ``simulated users'' complementing the design and evaluation process of interactive systems while reducing reliance on costly experimental testing~\cite{oulasvirtaComputationalRationalityTheory2022}.

Still, before models of visual sampling are able to exploit their capabilities to support research and design, four key challenges need to be addressed. They must be able to (1) Predict full eye movement patterns, enabling the analysis of gaze behavior; (2) Operate directly on raw pixel input, removing the need for handcrafted task representations; (3) Produce accurate predictions in fast-changing, dynamic environments where attentional demands shift rapidly; (4) Be grounded in theoretical models of perception and cognition, ensuring interpretability and supporting the modeling of individual differences among users.
To date, most existing models fall short on at least one of these key challenges. Many focus primarily on visual attention allocation in static settings, such as data visualizations~\cite{shi_chartist_2025}, single-screen app layouts~\cite{xu_spatio-temporal_2016, deza_attention_2017}, web pages~\cite{drusch_analysing_2020}, or natural scenes~\cite{coutrot_scanpath_2018, yang_predicting_2020}. Others rely on task-specific abstractions to manage the complexity of dynamic environments~\cite{yoon_modeling_2025}, limiting their generalizability.
Moreover, many of these models are black-box systems based on deep neural networks~\cite{lopez-cardona_comparative_2025}, which lack interpretability and offer limited insight into the cognitive mechanisms that drive human behavior. While they may predict where users are likely to look, they fail to explain why those decisions are made or how behavior might adapt to individual differences or changes in task demands. As a result, such models are of limited usage to answering ``what if'' questions that drive the design of interactive technology.

To fill this gap, we present CR-Eyes, a model based on the theory of \ac{CR}~\cite{gershmanComputationalRationalityConverging2015, oulasvirtaComputationalRationalityTheory2022} that simulates human sampling behavior in fast-paced, visually complex Atari gameplay.
In contrast to related approaches~\cite{yoon_modeling_2025} CR-Eyes works directly on raw pixel input and does not require hand-crafted task environments. We evaluate CR-Eyes on three Atari games using the Atari-HEAD dataset~\cite{zhangAtariHEADAtariHuman2019}.

Our results show that the model learns to sample the environment in a way that achieves game scores comparable to non-expert humans. While saliency maps derived from the agent’s behavior closely align with data generated by humans, the underlying eye-movement patterns still deviate substantially from human scanpaths.

\section{Related Work}
Traditionally, cognitive architectures such as EPIC \cite{kierasOverviewEPICArchitecture1997} and ACT-R \cite{salvucciIntegratedModelEye2001} have been used to model human attention allocation. While successful in \acs{HCI} tasks such as goal-directed search \cite{kierasVisualSearchSelective2019, halversonComputationalModelActive2011} and menu selection \cite{byrneACTRPMMenu2001}, these approaches rely on handcrafted environments and production rules, rendering them impractical for real-world applications.

A different line of work comes from robotics-inspired active vision systems, which learn visual sampling strategies to optimize task performance using sensory–motor reinforcement learning \cite{whiteheadActivePerceptionReinforcement1990}. For example, \citet{shang_active_2023} propose an active vision agent for Atari games in the \ac{ALE} \cite{bellemare_arcade_2013} with separate sensory and motor action heads. However, such systems do not incorporate human perceptual or cognitive constraints and therefore do not produce human-like gaze behavior or strategies.

One approach for modeling human-like behavior, combining recent developments in reinforcement learning and cognitive theory, is \acf{CR} \cite{oulasvirtaComputationalRationalityTheory2022}. CR frames human behavior as rational optimization under constraints imposed by the task design, the user’s perceptual, motor, and cognitive capabilities, and competing goals such as speed and accuracy. By explicitly modeling such bounds, these models are more interpretable than deep neural networks, because they learn an optimization under explicitly defined, human-like constraints. Furthermore, CR models allow to answer ``what if'' questions through adjustment of these constraints.
Recent examples of CR-based models that simulate human vision are Chartist that models top-down eye movements over static scenes \cite{shi_chartist_2025}, or Typoist that captures hand–eye coordination during touchscreen typing~\cite{shi2025typoist}. While Chartist~\cite{shi_chartist_2025} relies on a Large Language Model for high-level information acquisition, memory, and planning, to inform oculomotor control via RL on a pixel basis, Typoist~\cite{shi2025typoist} on the other hand, relies on screen-based raw pixel input for the RL agent alone. These models assume gaze-contingent perception in largely static or user-controlled environments. \citet{yoon_modeling_2025} extend this paradigm to a dynamic first-person shooter by modeling saccadic latency and peripheral uncertainty, but rely on structured state representations rather than raw pixels. In contrast, CR-Eyes operates directly on pixel input from a continuous video game and accounts for environment changing actions. 

To evaluate our approach we employ Atari-HEAD~\cite{zhangAtariHEADAtariHuman2019}, a large-scale dataset that captures over 117 hours of gameplay across 20 Atari 2600 games. In contrast to regular Atari games, the games in Atari-HEAD run at a lower frame rate of $20\,\mathrm{Hz}$ and could be completely paused, resulting in near-optimal information acquisition behavior and thus game performance, which we mirror to match the human action space and timing conditions during data collection.

\section{CR-Eyes}

\begin{figure*}
    \centering
    \includegraphics[width=0.7\linewidth]{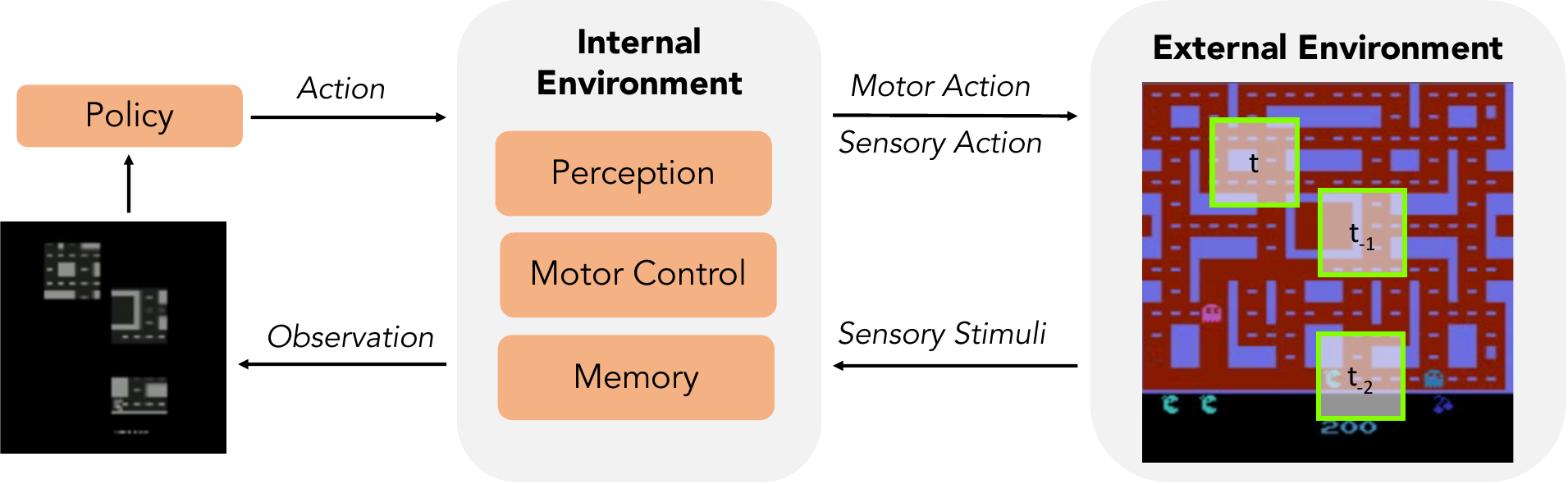}
    \caption{An overview over the CR-Eyes Architecture. Through the internal environment the agent performs a motor action (game action) and a sensory action (position to look at) at every timestep $t$ in the external environment. The resulting observed patch is returned to the internal environment and integrated into the memory, consisting of the last $n$ stacked observations. The memory is given to the RL agent as observation on which basis the agent will output the next motor and sensory action.}
    \Description{An architecture diagram consisting of the parts, the external environment, the internal environment, and the policy. The external shows an Atari game screenshot with three different gaze positions of the current and the two previous time steps. The internal environment shows the modeled bounds imposed by CR, namely perception, motor control, and memory. An arrow points from the internal environment to a black square with the same size as Atari screen from the external environment showing the same observed patches as in the external environment. This aggregated memory is given as observation to the policy. Arrow from Policy to Internal Environment with label "Action". Arrow from Internal Environment to External Environment with labels Motor Action and Sensory Action. Arrow from External Environment to Internal Environment with label Sensory Stimuli.}
    \label{fig:system_overview}
\end{figure*}

The goal of CR-Eyes is to reproduce human eye movements and gameplay behavior in Atari games while directly interacting with the \ac{ALE}~\cite{bellemare_arcade_2013} based on raw pixel inputs. To achieve these goals, we make the following design choices: (1) We leverage the theory of \acf{CR}, modeling humans' internal environment (e.g, perception, cognition, motor control) and their external environment (e.g., game design);
(2) In line with evidence from cognitive science~\cite{fodor1983modularity} and related work~\cite{shiCRTypistSimulatingTouchscreen2024, shi2025typoist}, we design distinct components to model perception, motor control and memory;
(3) Inspired by findings on human sensorimotor learning~\cite{wolpert2011principles}, we use \ac{SUGARL} as proposed by \citet{shang_active_2023} to learn joint motor and sensory policies. The whole architecture is shown  in Fig.~\ref{fig:system_overview}.

\subsection{Problem Formulation}
Following related work on \ac{CR} and visual sampling~\cite{shi_chartist_2025, shi2025typoist}, we formulate human visual sampling and gameplay as a sequential decision-making problem under uncertainty via a constrained, computationally rational \ac{POMDP} $(S, A, O, r)$:

 \begin{itemize}
     \item $S$ is the state space, where each state $s_t$ is defined by the pixel representation of the environment at time $t$.
     \item $A$ is the action space, in which the agent can execute a motor and a sensory action simultaneously. \emph{Motor actions} describe game actions, such as \emph{up, down, left, right,} or \emph{fire} or to pause the game at the current frame. \emph{Sensory actions} move the fovea, meaning they determine which portions of the screen to focus on.
     \item $O$ denotes the observation space, in which $o_t$ is represented by a stacked pixel representation of the last $n$ observed image patches.
     \item $r$ is the reward provided after every time step $t$. We formulate it as the sum of the achieved game score in a specific time step, e.g., the points dropped by the game for collecting objects and negative penalties for pausing and larger eye movements. 
     
 \end{itemize}

\subsection{External Environment}
The external environment is a semi-frame-by-frame environment adopted from the \ac{ALE}~\cite{bellemare_arcade_2013} framework. Accordingly, it adds the possibility to pause the game at each step and thereby mirrors the setup as used by \citet{zhangAtariHEADAtariHuman2019}, in particular a frame rate of $20\,\mathrm{Hz}$. The agent takes two actions every timestep, one sensory and one motor action. 
To determine the time passed between two steps, we use the EMMA model which allows us to calculate the time it took for the agent to fixate two subsequent patches of the screen. 

\subsection{Internal Environment}
Vision is the main interface between the agent and the game. Human vision is limited by several factors such as muscular bounds that result in longer saccade times for larger eye movements, or fixation and encoding times of targets. To model realistic saccade times, we employ the EMMA model by~\citet{salvucciIntegratedModelEye2001} with default parameters and replicating the setup by~\citet{zhangAtariHEADAtariHuman2019}. The internal environment models foveated vision, so that the agent only perceives a fraction of the screen displaying the Atari game, similar to \citet{shang_active_2023}, where the screen is divided into a $5 \times 5$ grid. These 25 discrete positions define the sensory actions available to the agent. Furthermore, we stack the last $n=4$ observed patches to model working memory that serves as the agent's observation.

\subsection{Neural Network Architecture and Training}
Based on the work of \citet{shang_active_2023} we utilize a similar architecture, namely a DQN with two separate action heads. Following best practices in Atari RL we re-scale the gameplay frames to $84\times84$ pixels, use sticky actions with $\xi=0.25$, and frame skipping $k=4$~\cite{machadoRevisitingArcadeLearning2018}.
To prevent indefinite pausing, we employ \emph{invalid action masking}~\cite{huangCloserLookInvalid2022} and introduce a specific pause penalty. This penalty, along with saccade costs, is optimized via hyperparameter grid search to match human data. 

\section{Evaluation and Results}
To evaluate CR-Eyes, we compare gameplay performance and gaze behavior against human benchmarks and datasets.
We evaluate CR-Eyes on three Atari games, namely \emph{Asterix, H.E.R.O.}, and \emph{Seaquest}. For these games, eye-tracking data~\cite{zhangAtariHEADAtariHuman2019} as well as non-expert human scores are available~\cite{bellemare_arcade_2013}.

\subsection{Gameplay Behavior}
We report game scores in both the standard setting ~\cite{bellemare_arcade_2013} and the pausing enabled Atari-HEAD setting~\cite{zhangAtariHEADAtariHuman2019} (15 minutes scores), testing whether CR-Eyes can leverage pausing to acquire additional information and improve action selection, similar to humans.
While the goal of CR-Eyes is to achieve human-like scores, we also compare our results against classical \ac{RL} approaches such as 
DQN~\cite{mnihHumanlevelControlDeep2015} and the world model approach Dream-er~\cite{hafnerMasteringDiverseDomains2023}.
Table~\ref{tab:scores} shows that CR-Eyes achieves game scores similar to non-expert humans in \emph{Asterix} and \emph{H.E.R.O.} for the regular game setting.
However, for \emph{Seaquest} CR-Eyes' scores are lower. 
While CR-Eyes achieves better game scores in all three games when it's given the possibility to pause the gameplay, the improvement is significantly lower than that observed in human players. The scores reported in the Atari-HEAD dataset~\cite{zhangAtariHEADAtariHuman2019} are several orders of magnitude higher than the best CR-Eyes scores. 
The DQN with full state knowledge also achieves scores levels of magnitude lower~\cite{mnihHumanlevelControlDeep2015} than those reported in Atari-HEAD~\cite{zhangAtariHEADAtariHuman2019}.  In contrast \citet{hafnerMasteringDiverseDomains2023} report that their approach \emph{Dreamer} achieves scores much higher than those reported in Atari-HEAD.

\begin{figure*}
    \centering
    \centering
    \includegraphics[width=0.7\linewidth]{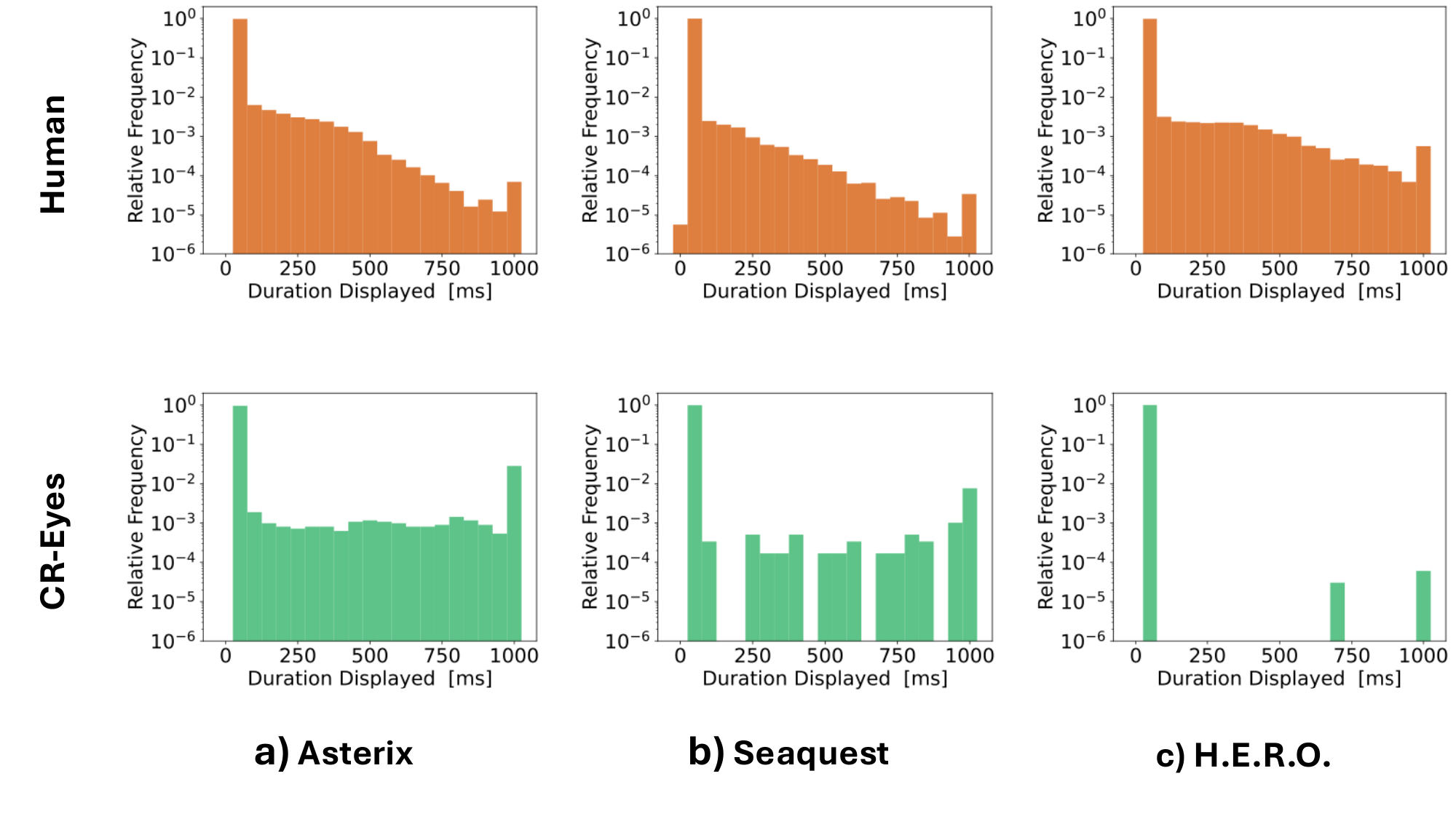}

    \caption[Duration frames were displayed for humans and our agent.]{Durations how long frames were displayed in \si{\milli\second} for humans (orange)~\cite{zhangAtariHEADAtariHuman2019} and our agent (green). The agent was trained for three million training steps. The long tail of the distribution is summed up in the last bin and the y-axis is logarithmically scaled.}
    \label{fig:durations} 
\end{figure*}

The \emph{pause frequency} is inferred by analyzing the time a frame was displayed (\autoref{fig:durations}). The human data (\autoref{fig:durations}~(top)) was recorded by~\citet{zhangAtariHEADAtariHuman2019} and the durations for CR-Eyes (Fig.~\ref{fig:durations}~(bottom)) are calculated using the EMMA model~\cite{salvucciIntegratedModelEye2001}. Notably, the game remains unpaused for the majority of the time for both human and CR-Eyes ($50ms$ display duration). Analyzing the deviations shows that humans prefer shorter pauses. In contrast, the agent has a roughly equal distribution of pause lengths. In general, the agent replicates the human trend to only pause the game seldom. However, for H.E.R.O. the agent learned to not pause at all.
Qualitative evaluations of gameplay videos show that, while humans mainly pause in complex situations, the agent primarily pauses shortly before losing the game. 

\begin{table*}[htb]
    \centering
    \small
    \setlength{\aboverulesep}{0pt}
    \setlength{\belowrulesep}{0pt}
    \renewcommand{\arraystretch}{1.2}

    \caption[Game performance of the CR Agent.]{Game scores for different human (orange) and RL (blue) baselines and our agent (green).}
    
    \begin{tabular}{
        c 
        >{\columncolor{pastelorange}}c 
        >{\columncolor{pastelorange}}c 
        >{\columncolor{pastelblue}}c 
        >{\columncolor{pastelblue}}c 
        >{\columncolor{pastelgreen}}c 
        >{\columncolor{pastelgreen}}c 
    }
        \toprule
        
        \makecell[c]{\textbf{Game}} & 
        \makecell[c]{\textbf{non-expert} \\ \textbf{Human} \cite{bellemare_arcade_2013}}  & 
        \makecell[c]{ \textbf{Atari-HEAD} \\ \cite{zhangAtariHEADAtariHuman2019}} & 
        \makecell[c]{\textbf{DQN} \\ \cite{mnihHumanlevelControlDeep2015}} & 
        \makecell[c]{\textbf{Dreamer} \\ \cite{hafnerMasteringDiverseDomains2023}} & 
        \makecell[c]{\textbf{CR-Eyes} \\  \textbf{(no~pause)}} &
        \makecell[c]{\textbf{CR-Eyes} \\ \textbf{(pausing)}} \\
        
        \midrule
        Asterix & 620 & 135,000 & 6012 & 441,763 &  495 & 619  \\
        Seaquest & 156 & 64,710 & 5286 & 356,584 & 20 & 68 \\
        H.E.R.O. & 6087 & 77,185 & 19,950 & 40,677 & 5112  & 8649 \\
        \bottomrule
    \end{tabular}
    \label{tab:scores}
\end{table*}

\subsection{Gaze Behavior}
\begin{figure*}
    \centering
    \begin{minipage}{0.55\textwidth}
        \centering
        
        \begin{subfigure}[t]{\wthree}
            \includegraphics[width=\textwidth]{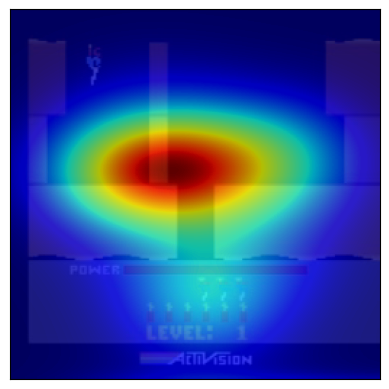}
            \caption{Human}
            \label{fig:episode-heatmap-human}
        \end{subfigure}
        \begin{subfigure}[t]{\wthree}
            \includegraphics[width=\textwidth]{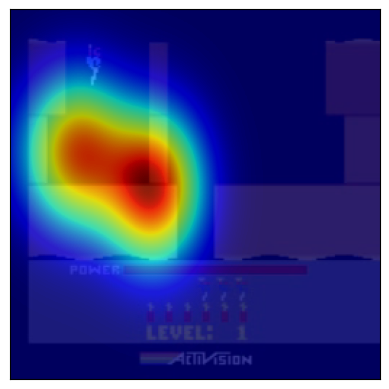}
            \caption{Agent}
            \label{fig:episode-heatmap-agent}
        \end{subfigure}
        \begin{subfigure}[t]{\wthree}
            \includegraphics[width=\textwidth]{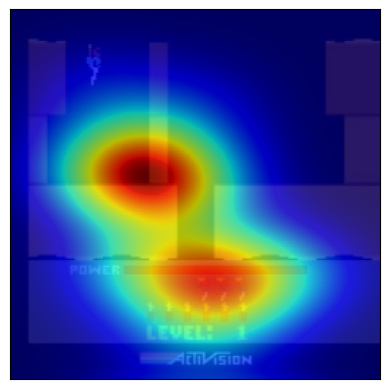}
            \caption{Truncated Human}
            \label{fig:episode-heatmap-short}
        \end{subfigure}

        \label{fig:episode-heatmap}
    \end{minipage}
    \Description{Three heatmaps comparing visual attention . (a) 'Human' displays a broad, central focus area. (b) 'Agent' shows a tighter, more concentrated focus shifted slightly to the upper left. (c) 'Truncated Human' exhibits a split focus with two distinct hotspots: one on the central gameplay area and a second on the lower half of the image.}
    \caption[Agent and human saliency throughout an entire episode.]{Agent and human saliency throughout an entire \emph{Seaquest} episode. Human data is taken from Atari-HEAD. Subfigure c) shows saliency for the human episode truncated to the length of the agent's episode.}
\end{figure*}

We compare CR-Eyes saliency predictions to those of human players in the Atari-HEAD dataset~\cite{zhangAtariHEADAtariHuman2019}.
Due to the stochastic environment and the environment changing actions, the game roll-outs of CR-Eyes do not correspond to the ones in the dataset. Accordingly, we cannot directly compare the saliency maps generated by CR-Eyes to those available in the dataset.
To solve this issue, we train a supervised saliency predictor using the Atari-HEAD data to generate synthetic saliency maps for the roll outs generated by CR-Eyes. Saliency prediction is a common technique used in RL and imitation learning to evaluate an agent's visual attention allocation against that of humans~\cite{mnihHumanlevelControlDeep2015, zhangAtariHEADAtariHuman2019, le_meur_methods_2013}.
Saliency maps over the whole episode reveal that unlike humans who mostly focus at the center (Fig.~\ref{fig:episode-heatmap-human}), the agent fixates on the upper-left spawn point (Fig.~\ref{fig:episode-heatmap-agent}), missing the next key \acp{AOI} relevant for game progress.  

\begin{figure*}
    \centering
    \includegraphics[width=0.9\linewidth]{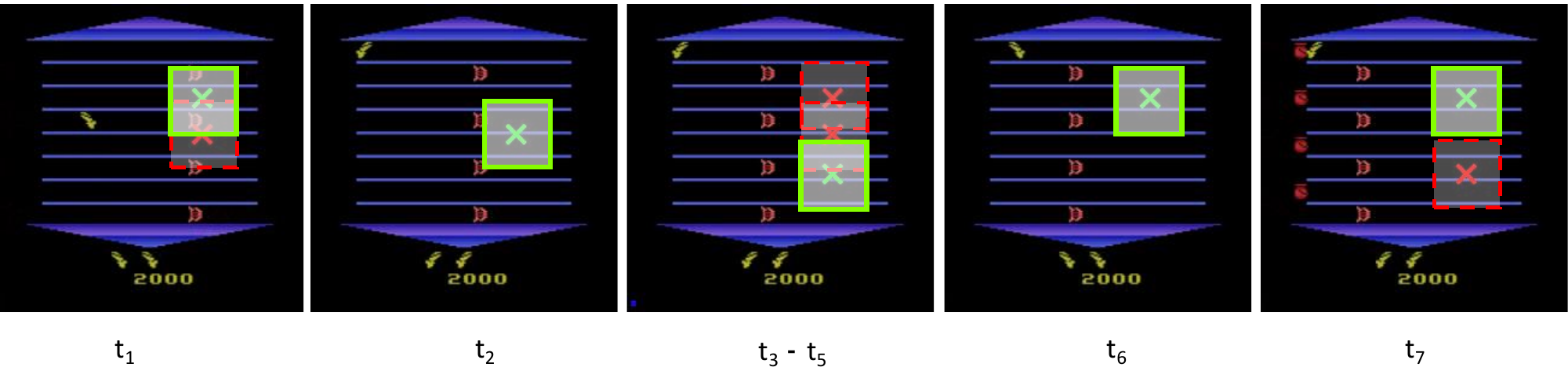}
    \caption{Resulting scanpath for \emph{Asterix}. The green squares are the fovea position at step $t_i$. The red squares represent the agent's sensory action, that will be executed in the next step. At $t_3$ the game is paused until $t_5$.}
    \Description{Five consecutive frames of Asterix showing the agent's foveal position moving from the upper right in t1 to the lower right in t5 and going back to the upper right in t6 and t7.}
    \label{fig:scanpath}
\end{figure*}

However, truncating human episodes to the mean length of agent episodes  (Fig.~\ref{fig:episode-heatmap-short}) reveals systematic differences in saliency. 
While the agent identifies the avatar as the relevant object on the screen, it fails to translate this knowledge into effective gameplay.
This behavior can also be observed in the scanpaths as shown in \autoref{fig:scanpath}. In this scene the avatar is in the top left corner, but the agent focus remains in the middle right and empty part of the screen.

\section{Discussion and Future Work}
Our results show that CR-Eyes is a first step toward a computational model grounded in cognitive science that can simulate human gaze behavior by closing the perception-action loop in dynamic environments solely based on pixel streams. To evaluate if CR-Eyes simulates human behavior, we compare it to human data along four dimensions: (1) game scores, (2) saliency predictions, (3) scanpaths, and (4) pausing behavior.
CR-Eyes's game performance is comparable to non-expert humans in a regular setting without pausing.
Like humans, CR-Eyes leverages the possibility to pause the gameplay to reach higher game scores. However, this improvement is magnitudes smaller compared to that of human players. Furthermore, our approach does not reflect human preferences for short pauses.
We further compared saliency predictions against human data from Atari-HEAD~\cite{zhangAtariHEADAtariHuman2019}. While CR-Eyes aligns closely to human saliency in early game stages it slightly diverts in later stages.
Simulated scanpaths however do not match human data. The agent does not focus on the avatar and regularly misses information considered relevant by humans such as approaching enemies.

From a methodological perspective, these results suggest that saliency maps alone are an insufficient metric for evaluating visual sampling behavior as saliency maps can mask fundamentally different sampling behavior.
From a behavioral perspective, the agent’s scanpaths  are primarily optimized for reward maximization, whereas human visual behavior reflects an understanding of task dynamics, leading to anticipatory glances based on expectations rather than immediate information gain. This indicates that the agent does not acquire game dynamics in a human-like manner. Differences in pausing behavior further suggest that the agent fails to integrate multiple observations to better infer and exploit environment dynamics.
This limitation likely stems from the model-free nature of DQN. Even in standard fully observable Atari environments, DQN scores magnitudes lower than humans in Atari-HEAD. To bridge this gap, model-based algorithms like Dreamer~\cite{hafnerMasteringDiverseDomains2023} appear promising, as they learn world dynamics to facilitate strategic planning. 
Another reason for the observed differences in visual sampling behavior may be that the agent relies on foveated vision only, whereas humans use peripheral vision as well. Research shows that this affects performance and may result in different gaze patterns and scanpaths~\cite{eisma_expectancy_2024}.

We propose the following steps for future work: (1) use a model-based algorithm like Dreamer~\cite{hafnerMasteringDiverseDomains2023} to allow the agent to generate an internal representation of game dynamics, (2) perform parameter inference with human behavior as cost function, and (3) employ models of peripheral vision~\cite{rosenholtzCapabilitiesLimitationsPeripheral2016}, bringing agent perception closer to human perception.
While gaming behavior is achievement-oriented, it shares the core cognitive challenge of information acquisition in dynamic environments with high-stakes domains. We argue that the gap between entertainment and safety critical tasks is defined by the cost of failure. As CR models explicitly optimize against a utility function, this can be bridged by adjusting the penalty for errors to reflect high-stakes consequences. Thus, our model is a first prototype and could be further applied to other domains such as automated testing of interface designs in driving~\cite{lorenz2024computational}, aviation, or video game design, complementing user studies. Additionally, in high-stake monitoring tasks like air traffic control~\cite{gasse_evaluating_2025}, these models could serve as a safety mechanism by detecting irregular scanning behavior and preventing \emph{Out-of-the-Loop}~\cite{merat_out_loop_2019} effects.

\begin{acks}
The authors acknowledge the financial support by the Federal Ministry of Research, Technology and Space of Germany and by Sächsische Staatsministerium für Wissenschaft, Kultur und Tourismus in the programme Center of Excellence for AI-research „Center for Scalable Data Analytics and Artificial Intelligence Dresden/Leipzig“, project identification number: ScaDS.AI.
\end{acks}

\balance
\bibliographystyle{ACM-Reference-Format}
\bibliography{references}

\clearpage

\end{document}